# An electric vehicle charging station access equilibrium model with M/D/C queueing

**Bingqing Liu, Theodoros P. Pantelidis, Stephanie Tam, Joseph Y. J. Chow***

C2SMART University Transportation Center, Department of Civil and Urban Engineering, New York University, New York, NY, USA

*Corresponding author's email: joseph.chow@nyu.edu

**ABSTRACT**

Despite the dependency of electric vehicle (EV) fleets on charging station availability, charging infrastructure remains limited in many cities. Three contributions are made. First, we propose an EV-to-charging station user equilibrium (UE) assignment model with a M/D/C queue approximation as a nondifferentiable nonlinear program. Second, to address the non-differentiability of the queue delay function, we propose an original solution algorithm based on the derivative-free Method of Successive Averages. Computational tests with a toy network show that the model converges to a UE. A working code in Python is provided free on Github with detailed test cases. Third, the model is applied to the large-scale case study of New York City Department of Citywide Administrative Services (NYC DCAS) fleet and EV charging station configuration as of July 8, 2020, which includes unique, real data for 563 Level 2 chargers and 4 Direct Current Fast Chargers (DCFCs) and 1484 EVs distributed over 512 Traffic Analysis Zones. The arrival rates of the assignment model are calibrated in the base scenario to fit an observed average utilization ratio of 7.6% in NYC. The model is then applied to compare charging station investment policies of DCFCs to Level 2 charging stations based on two alternative criteria. Results suggest a policy based on selecting locations with high utilization ratio instead of with high queue delay.

# 1 INTRODUCTION

Electric vehicles (EVs) are critical to reducing $CO_2$ emissions, improving air quality and promoting the usage of alternative fuels. The technological breakthroughs in the field (increased driving range, reduced charging time) have ignited the demand for EVs and by 2040, 57% of all passenger vehicles are projected to be electric worldwide (Valdes-Dapena, 2019). In New York City (NYC), for example, the 80 × 50 Roadmap (NYC, 2016) recommends investing in the largest city-owned light-duty electric vehicle fleet in the U.S. (set to 4000 vehicles by 2025 in 2019 (Government Fleet, 2019)) to help achieve 80% greenhouse gas (GHG) emissions reduction by 2050.

On average, it takes 20 to 30 minutes for Direct Current Fast Chargers (DCFCs) and 4 to 6 hours for Level 2 chargers to fill up an 80-mile battery (Chargepoint, 2020). Such a long charging time makes the viability of electric vehicles highly dependent on the availability of charging infrastructure (Jung et al., 2014; Jung and Chow, 2019). More specifically, the positioning of EV charging stations influences users' experience. For example, studies have shown that users tend to prefer charging stations located closer to them to reduce delay (Sun et al., 2016), although the long charging time can also induce long queues (Jung et al., 2014; Chen et al. 2017; Pantelidis et al., 2021).

Despite the dependency of EV fleets on charging station availability, the numbers of charging stations in many cities remain limited. For example, there were only 4 public DCFC stations available to the NYC-employed fleet in July 2020, where the rest of the 263 charging stations at that time were Level 2 chargers (NYC DCAS, 2020). The city has plans to invest in more charging infrastructure (NYC, 2020).

Given the cost and importance of charging station positioning with consideration of queue delay, an EV-to-charging station assignment model with proper charging congestion is needed to evaluate investment plans (e.g. Berger, 2021). With the assignment model, a configuration can be evaluated by changes in such metrics as access time, queue delay, utilization ratio, and charging time. To develop such a model, a queueing model is needed to describe the queueing systems at charging stations given the charging times and capacities (Jung et al., 2014). However, the literature on vehicle-to-charging station assignment is limited, and these limited studies rarely apply their proposed models to real world data, testing them only on toy networks like Sioux Falls or Nguyen-Dupuis networks. Much research has focused instead on facility location with assignment that ignores queueing (e.g. Upchurch and Kuby, 2010), adding route choice to the charging station choice as part of a general traffic assignment (He et al., 2014). There are some efforts on modeling the station queueing, including static approximations like the Davidson's cost functions (Zhou et al., 2018), as well as queueing models such as M/M/C (Said et al., 2013), M/M/1/K (Biondi et al., 2016), M/M/C/K (Yang et al., 2017), and M/G/1 (Keskin et al., 2019) queues.

Assuming M/M/C or M/M/C/K queues result in unrealistic charging time distributions. Exponentially distributed charging times give nonnegligible probabilities for unrealistically high charging times under any parameter settings. For example, if a charger with a 3-hour average charging time is modeled with an M/M/1 queue, the probability of 5-hour, 6-hour, and 7-hour charging time would be 18.9%, 13.5%, and 9.7%, even if the maximum charging time is 4 hours. This is unrealistic and leads to overestimated delays. Based on the above research gap, we seek to develop an EV-to-charging station assignment model with more realistic charging station congestion modeling, which could be efficiently used as an evaluation tool of city-scale charging station configurations. The tool should be able to model the UE under certain geographical EV distribution as well as the charging station configuration which is being evaluated.

In this study, an EV-to-charging station assignment model with M/D/C charging station queueing is proposed to evaluate different EV charging station location configurations to plan for future expansion. A new EV assignment model is proposed as a UE problem, originally introduced in Beckman et al. (1956), to capture users' charging station choices. The limited capacity of charging station is modeled by the M/D/C queueing model, which assumes users arrive randomly while charging time is deterministic (e.g. we assume a dead 80-mile battery will take half hour to fully recharge with a DCFC). Since there is no closed-form expression for an M/D/C queueing system, we use an approximation from the literature (Barceló et al., 1996). The model is solved using the Method of Successive Averages (MSA) which guarantees convergence for convex problems (Powell and Sheffi, 1982) like the EV-to-charging station assignment. We assume that users seek public charging stations close to home (Hardman et al., 2018) while avoiding queue delay.

We implement this model using a real, large-scale dataset of zonal-aggregated EV user home locations and charging stations as of July 8, 2020, shared by the New York City Department of Citywide Administrative Services (NYC DCAS). Based on the calibrated assignment results of the current configuration, we further demonstrate its applicability to policy analysis by evaluating 10 different improvement policies, using evaluation metrics including system total/average access time and charging time. In particular, we compare growth strategies for investing DCFCs at existing Level 2 charging stations: (1) those with the highest utilization ratios versus (2) those with the highest queue delays. We conclude that the policy of upgrading Level 2 charging stations with high utilization ratios deserves more attention than the one selecting high queue delay locations. These insights will prove useful to policymakers as New York (NY) builds out their charging infrastructure (Berger, 2021; NY, 2021).

The study is organized as follows. Section 2 presents the literature review. Section 3 presents the model formulation and the evaluation of different charging station congestion parameters. Section 4 presents the NYC case study. Section 5 provides a conclusion this study as well as future research directions.

## 2 LITERATURE REVIEW

Table 1 provides an overview of recent literature on EV charging station assignment models. The EV-to-charging station assignment problem has been studied quite extensively, typically within the context of a facility location problem for determining charging station locations. Many of these studies considered locations that satisfy route-level demand as a type of flow interception problem (see Upchurch and Kuby, 2010), i.e. travelers seek shortest paths in a congested network for their trip such that they may have to deviate to recharge along the way. A user equilibrium (or, in some cases, a stochastic user equilibrium) is used to capture the steady state behavior of the EV drivers (He et al., 2014; Lee et al., 2014; Riemann et al., 2015; Jing et al., 2017; He et al., 2018; Zhou et al., 2018). However, with typical commutes being less than an hour long and charging activities that can take half an hour or several hours, a prevalent design objective should be to determine charging station locations to target users who prefer to charge near their homes instead (Hardman et al., 2018).

**Table 1**. Recent literature of EV-to-charging station assignment models

| Authors | Assignment structure | Assignment solution method | Station congestion |
|---|---|---|---|
| Upchurch & Kuby (2010) | Flow interception access | Greedy heuristic for FRLM | N/A |
| Jung et al. (2014) | Queueing based on arrivals generated from taxi simulation | Simulation-based: vehicle choose closest charging station | M/M/C |
| He et al. (2014) | Route choice equilibrium with charging constraint | Iterative solution (path generation) | N/A |
| Lee et al. (2014) | Flow interception (charge ratio) with UE route choice constraint | Locations solved by enumeration with Frank-Wolfe for UE | N/A |
| Riemann et al. (2015) | Flow interception with stochastic user equilibrium (SUE) route choice constraint | Linearization and CPLEX | N/A |
| Biondi et al. (2016) | Set covering location problem | Greedy algorithm | M/M/1/K |
| Jing et al. (2017) | Set covering with SUE route choice constraints | Iterative heuristic for bi-level problem | N/A |
| Yang et al. (2017) | Charger allocation problem | Linearization and Gurobi for ILP | M/M/C/K |
| He et al. (2018) | Bi-level station location problem with UE route choice constraint | CPLEX with Frank-Wolfe algorithm for UE assignment | N/A |
| Zhou et al. (2018) | Route choice NLP with SUE route choice constraint | Frank-Wolfe algorithm | Davidson function |
| Keskin et al. (2019) | EVPTW with queue delay at charging stations | Adaptive Large Neighborhood search | M/G/1 |
| *This study* | *Charging station choice equilibrium with M/D/C queueing* | *MSA* | *M/D/C* |

Another area of research is on adding queue delay to the evaluation of charging station locations, again typically within a facility location context (exceptions include vehicle routing problem in Keskin et al. (2019)). Jung et al. (2014) was one of the first to explicitly model queueing within a charging station location problem as a simulation-based optimization problem (to deal with electric taxi itineraries). Their model considers an M/M/C queue approximation, which is generally regarded as overly simplistic and unrealistic. Later studies considered other forms of queues: M/M/1/K (Biondi et al., 2016), M/M/C/K (Yang et al., 2017, ), Davidson function to approximate M/M/1 (Zhou et al., 2018), and M/G/1 (Keskin et al., 2019) queues. Some studies have also looked at queueing networks (e.g. Said et al., 2013; Tan et al., 2014). In these problems the demand is already assigned exogenously to each station so there is no allocation, or assignment, of demand to stations.

Table 1 points to a clear research gap: there is no equilibrium assignment model evaluating a set of charging station locations relative to EV home locations, and certainly none that consider more complex queue delay functions along with the vehicle assignment. M/M/C and M/M/C/K are the most used queueing models, while as discussed in section 1, they give nonnegligible probabilities for extremely large charging times which far exceeds the longest charging time needed. An M/D/C queueing model provides less random charging times, which avoids the probability of unrealistic Exponentially distributed charging times of M/M/C. Although M/D/C doesn't consider any randomness in charging times, which makes it unsatisfactory for microscopic charging behavior modeling, it is sufficiently accurate for macroscopic planning purpose (analogous to the use of deterministic user equilibrium in traffic planning despite random travel times and demand). Average charging time data is also easier to acquire, which make the assignment tool easier to use. However, it is more challenging to model M/D/C than a random M/M/C queue, since there is no closed-form expression for the expected queue delay of M/D/C queues. Several studies have proposed approximations of M/D/C queue waiting times (Rolfe, 1971; Cosmetatos, 1975; Barceló et al., 1996) and achieved high accuracy in estimating average wait times for medium and high server loads. The latest one (Barceló et al., 1996) is adopted for our equilibrium model.

## 3 EV CHARGING STATION ACCESS EQUILIBRIUM MODEL

Based on the research gap identified in the previous section, the research objective is to develop a model that forecasts the assignment of EVs to charging stations so that their location configuration can be evaluated taking queueing at charging stations into consideration. In this section, we define the EV-to-charging station assignment problem and propose the formulation of the assignment and queues. Notations are shown in Table 2.

**Table 2.** Notation

| | |
|---|---|
| $N$ | Set of nodes where the EV users are located. |
| $S$ | Set of nodes where the charging stations are located. |
| $\lambda_i$ | Number of EVs at within Traffic Analysis Zone (TAZ) $i$, where $i \in N$ needing to recharge per day. |
| $d_{ij}$ | Travel time (day) of the shortest path between node $i \in N$ and charging station $j \in S$. |
| $k_j$ | Number of chargers of the same type at charging station $j \in S$. |
| $\mu_j$ | Charging rate (veh/day) of chargers at charging station $j \in S$. This can be defined by class as well, where $S_m$ is the subset of charging stations of class $m$, and all service rates of $j \in S_m$ have charging rate $\mu_m$. |
| $W^D_{k_j}$ | Sum of expected queue delay and charging time (day/veh) of charging station j, which is assumed to be a M/D/k queue with $k_j$ servers. |
| $W^M_{k_j}$ | Sum of expected queue delay and charging time (day/veh) of a M/M/k queue with $k_j$ servers. |
| $W^D_{q,k_j}$ | Expected queue delay (day/veh) of charging station j, which is assumed to be a M/D/k queue with $k_j$ servers. |
| $W^M_{q,k_j}$ | Expected queue delay (day/veh) of a M/M/k queue with $k_j$ servers. |
| $\epsilon$ | Convergence criterion constant. |
| $X_{ij}$ | Indicates percent of total demand at node $i \in N$ assigned to a charging station $j \in S$. |
| $T_{ij}$ | Total cost for a user from node $i \in N$ recharging at station $j \in S$, including service time. |
| $\omega_{acc}$ | Weight of access time (sum of travel time and queue delay). |
| $\omega_{char}$ | Weight of charging time. |

## 3.1. Model formulation
*3.1.1 Problem Description*

The goal of the assignment model is to evaluate a certain EV charging station configuration under certain charging demand (EV home locations). More specifically, we seek to obtain a set of zone-based EV-to-charging station assignment flows such that no user can change their choice of charging station without being worse off in terms of a combination of travel time to the charging station, charging time, and queue delay due to waiting for an available charging station. With charging demand and corresponding charging stations given, the problem to which the solution is the EV-to-charging station assignment flows is what we define as the EV-to-charging station assignment problem.

Consider a set of nodes $N$ generating user demand for EV charging at an average rate of $\lambda_i$ ($veh/day$), $i \in N$ assumed to represent home locations. The set of charging stations is $S$. Each charging station $j \in S$ has $k_j$ chargers each with charging rate $\mu_j$. Travel time is a constant $d_{ij}$ $for$ $i \in N, j \in S$, which is pre-computed based on the shortest path and an average travel speed (25 mph). Each user from zone $i \in N$ recharging at a station in zone $j \in S$ incurs two delay components when recharging: expected queue delay and charging time $W_{k_j}^D$, and travel time $d_{ij}$. We assume deterministic charging time ($\frac{1}{\mu_j}$ for $j \in S$) for the same type of charger. Different types of chargers can have different charging times. Hence, the total time that a user spends from $i \in N$ to recharge at charging station $j \in S$ is represented by $W_{k_j}^D + d_{ij}$. We define *access time* as the sum of travel time $d_{ij}$ from node $i \in N$ to the charging station $j \in S$ and queue delay $W_{q,k_j}^D$ with $k_j$ servers. Considering that people may value access time and charging time differently, different weights can be assigned to access time ($\omega_{acc}$) and charging time ($\omega_{char}$) in the cost function. Hence, the total "cost" of a user from $i \in N$ to recharge at charging station $j \in S$ is $T_{ij} = \omega_{acc}\left(W_{k_j}^D + d_{ij} - \frac{1}{\mu_j}\right) + \frac{\omega_{char}}{\mu_j}$. The system total cost is defined as $T = \sum_{i \in N} \sum_{j \in S} \lambda_i X_{ij} T_{ij}$. Charging pricing is not considered in this study for the following two reasons. First, it costs approximately $2.4 to $3.6 to travel 25 miles using Level 2 chargers, while using DCFCs it costs $3.62 to travel the same distance (MYEV.com, 2021). This means the cost is approximately the same regardless of station type, hence we consider ignoring charging prices acceptable in this study. Second, in the case study, both charging supply and charging demand are owned by the city, which means that charging prices might not be a key factor in charging station choices. When modeling the assignment of private EVs to public charging stations, charging prices can be added to charging cost to construct a generalized charging cost function (all time costs transferred to dollars using value of time). We define this as a charging station access user equilibrium as shown in Definition 1, based on Wardrop's (1952) principle of user equilibrium for road traffic.

**Definition 1**. A system is in a **charging station access user equilibrium** if no user can unilaterally change their station choice without increasing their cost $T_{ij}$.

For this problem, the congestion effect (where the delay is dependent on the flow) only exists in the queues at the charging stations, i.e. we are not modeling road traffic congestion. The reason is that, as a small portion of traffic where the recharging activities may not even be at peak commuting hours or count for only a small increase (e.g. extra 15 minutes to access a nearby station locally) compared to the queue delay differential that can exist between of 0.5 hour and 4 hour charging times among limited numbers of DCFCs and Level 2 chargers. In addition, the purpose of this model is not intended for designing road infrastructure but for capacity planning for EV charging stations.

*3.1.2 Queueing Model*

With deterministic charging time, random arrival, and multiple chargers at one charging station, the M/D/C queueing discipline is assumed to model the queues at charging stations. We assume a first-come-first-serve queueing system with a buffer of infinite size with no limit to the number of customers in the queue (there is no "unmet demand"). The sum $W_C^D$ of expected queue delay $W_{q,C}^D$ and charging time $\frac{1}{\mu}$ is computed as Eq. (1). The queue delay represents the steady state delay of a dynamic system in which random, uncoordinated arrivals may lead to bunched or intermittent arrivals throughout the day. While the steady state expression is static, it considers such dynamics explicitly.

$$W_C^D = W_{q,C}^D + \frac{1}{\mu} \tag{1}$$

The average M/D/C queue delay $W_{q,C}^D$ does not have a closed form expression. As such, we present an approximation from Barceló et al. (1996), who used an M/M/C queue delay as shown in Eq. (2) – (4), where utilization ratio $\rho = \frac{\lambda}{\mu C}$, $C$ is the number of servers, and $\alpha_i$ is a recursive term used in determining the M/M/C queue delay. Adopting this approximation may lead to some inaccuracies, so we conduct extensive tests to verify its accuracy and identify the limitations. Like with continuous approximation and other parsimonious models in the literature, the objective is to build a tractable planning model to compare charging station locations at large scale.

$$W_{q,C}^D = \frac{W_{q,C}^M}{2}\left(1 + \frac{(1-\rho)(C-1)(\sqrt{4+5C}-2)}{16\rho C}\right) \tag{2}$$

$$\alpha_1 = 1, \alpha_i = 1 + \frac{\mu}{\lambda}(i-1)\alpha_{i-1} \tag{3}$$

$$W_{q,C}^M(\lambda,\mu) = \frac{\lambda}{(C\mu-\lambda)^2\left(\alpha_C + \frac{\lambda}{(C\mu-\lambda)}\right)} \tag{4}$$

To ensure that this approximation is sufficiently accurate, we simulate random arrivals that occur at a rate $\lambda$ with deterministic service times set to $\mu = 1.0$. Since the approximation from $W_{q,C}^M$ to $W_{q,C}^D$ is only dependent on $\rho$ and $C$, we assume fixed $\mu$ while varying $\rho$ and $C$. The simulated delays are then compared to the delays predicted by Eq. (2) – (4) to obtain absolute and relative errors. The results are shown in Figure 1, which are obtained for each combination of $(\rho, C)$ from a five-run average of simulated periods of 10000 seconds each. Generally, for all the $C$ values tested, absolute errors are low at low utilization ratios, and relative errors are low at high utilization ratios. The formula was shown to produce more accurate predictions for lower $C$ values. Overall, the M/D/C approximation illustrates quite satisfactory results for steady-state conditions for low values of $C$. For $C = 1$, the relative (%) errors only escalate up close 10% when $\rho$ is close to 0, which is a non-issue since the absolute error at low utilization ratios is negligible.

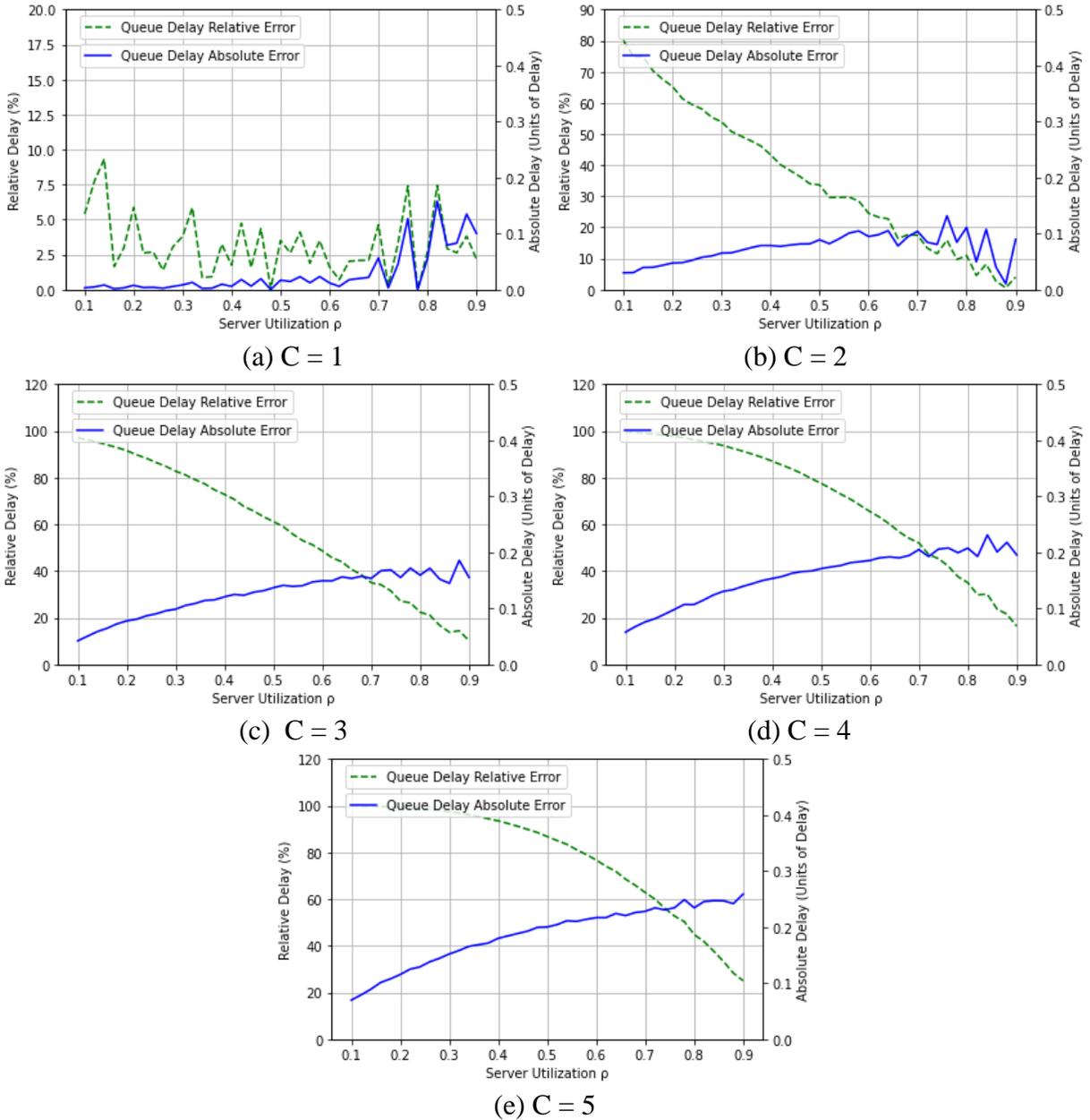

(a) C = 1

(b) C = 2

(c) C = 3

(d) C = 4

(e) C = 5

**Figure 1.** M/D/C Approximation Approach for Different C Values

*3.1.3 EV-to-Charging Station Assignment Model*
Having defined the equilibrium condition and the queue delay function and tested it, we present the nonlinear programming formulation for the charging station user equilibrium assignment model in Eq. (5) – (7). Objective (5) minimizes the sum of travel time and the integral of the sum of the queue delay and charging time at each station, where the queue delay is a measure of congestion at the charging station. The travel time can be defined as the time from the vehicle zones to the station, double that amount for roundtrips, or some accessibility-based measure in units of time that accounts for propensity for trip chains. In this study, we simply assume one-way travel access since we do not have trip chain data and do not know whether the vehicle would return home or go elsewhere. Due to the relationship between the charging station queue delay and the vehicles assigned to that station, the objective is non-linear and $W_{C_j}^D$ for station $j$ needs to be approximated by Eq. (1) – (4). The assignment constraint in Eq. (6) ensures all the demand nodes are matched to charging stations. Eq. (7) are the non-negativity constraints.

$$\min Z = \sum_{i \in N} \sum_{j \in S} \lambda_i d_{ij} X_{ij} + \sum_{j \in S} \int_0^{\sum_{i \in N} \lambda_i X_{ij}} W_{C_j}^D(x, \mu_j)\, dx \tag{5}$$

$$s.t.$$

$$\sum_{j \in S} X_{ij} = 1, \quad \forall i \in N \tag{6}$$

$$X_{ij} \geq 0, \quad \forall i \in N, j \in S \tag{7}$$

The model is structured similarly to the classic UE traffic assignment model. However, the structure of the $W_{C_j}^D$ term makes it difficult to both compute the value of $Z$ and to determine the derivative $\frac{\partial W_{C_j}^D(x, \mu_j)}{\partial \sum_{i \in N} \lambda_i X_{ij}}$ needed for derivative-based methods like Frank-Wolfe algorithm (see LeBlanc et al., 1975). Knowing that M/D/C queue delay functions are convex with respect to $\lambda$ and $c$ (Rolfe, 1971), we adopt a derivative-free algorithm to solve the model: Method of Successive Averages (MSA).

### 3.2. Proposed solution algorithm

MSA has been shown to be globally convergent (Powell and Sheffi, 1982) for convex problems and is commonly used to solve SUE traffic assignment models. The algorithm involves iteratively identifying an auxiliary solution that would optimize Eq. (5) as the search direction in the solution space, and then using MSA to determine the step size. Although Eq. (5) cannot be solved directly (due to unavailability of the integral expression). As such, we propose an original algorithm that uses the same shortest path logic from LeBlanc et al. (1975) to find an auxiliary assignment solution based on selecting the minimum costs at that iteration, and the MSA to find the step size. The algorithm is summarized as follows: 1) each iteration it updates auxiliary assignments based on the latest costs, 2) apply MSA to update the iteration's assignments, 3) check stopping conditions, and 4) update queue delays and total costs based on the iteration's assignment. While we will not be able to compute $Z$, the objective value is meaningless in any case and it is more useful to compute $T$ as the final performance measure.

**EV-to-charging station Assignment Algorithm: MSA for Charging Station Access Equilibrium**

**Input**: $N$: list of EV locations.
$S$: list of charging stations.
$\lambda$: $len(N)$ dimensional vector. $\lambda_i$ is the number of EVs at $i \in N$.

$k$: $len(S)$ dimensional vector. $k_j$ is the number of chargers at $j \in N$.
$\mu$: $len(S)$ dimensional vector. $\mu_j$ is the service rate of 1 charger at $j \in N$.
$d$: $len(N) \times len(S)$ dimensional matrix. $d_{ij}$ is the travel time between $i$ and $j$.
$\Delta$: tolerance

**Output**: $X^n, T^{n+1}$

**Initialize**: $T_{ij}^0 = d_{ij}$ for all $i \in N, j \in S$.
$X^0$, which is $len(N) \times len(S)$ dimentional matrix. $X^0[i,j] = 0$ for $i \in N, j \in S$.
$Y$, which is $len(N) \times len(S)$ dimentional matrix. $Y[i,j] = 0$ for $i \in N, j \in S$.
$\varepsilon = \infty$
$n = 0$

**while** $\varepsilon > \Delta$ **do**
    **for** all $i \in N$ **do**     //update auxiliary assignments
        $Y[i, \underset{j \in S}{\operatorname{argmin}}\{T_{ij}^n\}] \leftarrow 1$
    **end for**
    **if** n = 0 **then**     //MSA update of iteration assignment
        $X^0 \leftarrow Y$.
    **else**
        $X^n \leftarrow \frac{1}{n+1}Y + \frac{n}{n+1}X^{n-1}$
        $\varepsilon \leftarrow ||X^n - X^{n-1}||_2$
    **end if**
    **for** $j \in S$ **do**     //update queue delays based on latest assignment
        $A_j^n \leftarrow X[:,j] \cdot \lambda$
        $\rho_j^n \leftarrow \frac{A_j^n}{\mu_j k_j}$
        **if** $\rho_j^n < 1$ **then**
            **for** m = 1, 2, …, $k_j$ **do**
                **if** m = 1 **then**
                    $\alpha_m \leftarrow 1$
                **else**
                    $\alpha_m \leftarrow 1 + \frac{\mu_j}{A_j^n}(m-1)\alpha_{m-1}$
                **end if**
            **end for**
            $W_{q,k_j}^{M,n} \leftarrow \frac{A_j^n}{\left(\mu_j k_j - A_j^n\right)^2 \left(\alpha_{k_j} + \frac{A_j^n}{\left(\mu_j k_j - A_j^n\right)}\right)}$
            $W_{k_j}^{D,n} \leftarrow \frac{W_{q,k_j}^{M,n}}{2}\left(1 + \frac{(1-\rho_j^n)(k_j-1)(\sqrt{4+5k_j}-2)}{16\rho_j^n k_j}\right) + \frac{1}{\mu_j}$
        **else**
            $W_{k_j}^{D,n} \leftarrow \frac{A_j^n}{2\mu_j k_j} + \frac{1}{\mu_j}$

```
            end if
        end for
        for  i ∈ N do                     //update the total costs
            for  j ∈ S do
                    T_{ij}^{n+1} ← d_{ij} + W_{k_j}^{D,n}
            end for
        end for
        n ← n + 1
end while
```

The term $a_j^n(X)$ is used to capture the $\sum_{i \in N} \lambda_i X_{ij}$. In the "update auxiliary assignment" step, we assign all the EVs at each TAZ to the charging station with the smallest total cost $T_{ij}^n$ to find an auxiliary assignment vector $Y$. Then we compute the successive average of the assignment vector $Y$ at this iteration and all the $X$s at the previous $(n - 1)$ iterations to find a step size to update the assignment matrix $X^n$. Costs are then updated based on that assignment for the new iteration $n + 1$. This approach avoids computing Eq. (5) or the derivative $\frac{\partial W_{C_j}^D(x, \mu_j)}{\partial \sum_{i \in N} \lambda_i X_{ij}}$.

When computing queue delay, the formula does not handle $\rho > 1$. During intermediate iterations, some nodes may end up in such a state. For those cases, we compute a worst-case deterministic queue delay assuming all the EVs assigned to this charging station arrive at the same time. The queue delay of the last vehicle in the queue is computed as shown in Eq. (8), where $\mu, \lambda$, and $k$ denote the service rate, arrival rate, and number of chargers at the charging station.

$$W_k^D = \frac{\lambda}{2\mu k}, \quad \rho > 1 \tag{8}$$

The code is implemented and tested in Python and can be publicly accessed at our GitHub repository with instructions to apply to any study area: https://github.com/BUILTNYU/EV_Charging_Station_Access_Equlibrium_Model.

### 3.3. Illustrative case
We build a toy example with 10 EV parking locations and 5 charging stations to illustrate how the equilibrium model and the MSA algorithm works. The x, y coordinates of the EV parking locations and charging stations are shown in Figure 2 along with service rates for each charging station and EVs arrival rates at each parking location. We compute the Euclidean distances between each charging station and EV parking location to form an OD travel time matrix with the assumption that the average speed of EVs is 20-unit length/unit time.

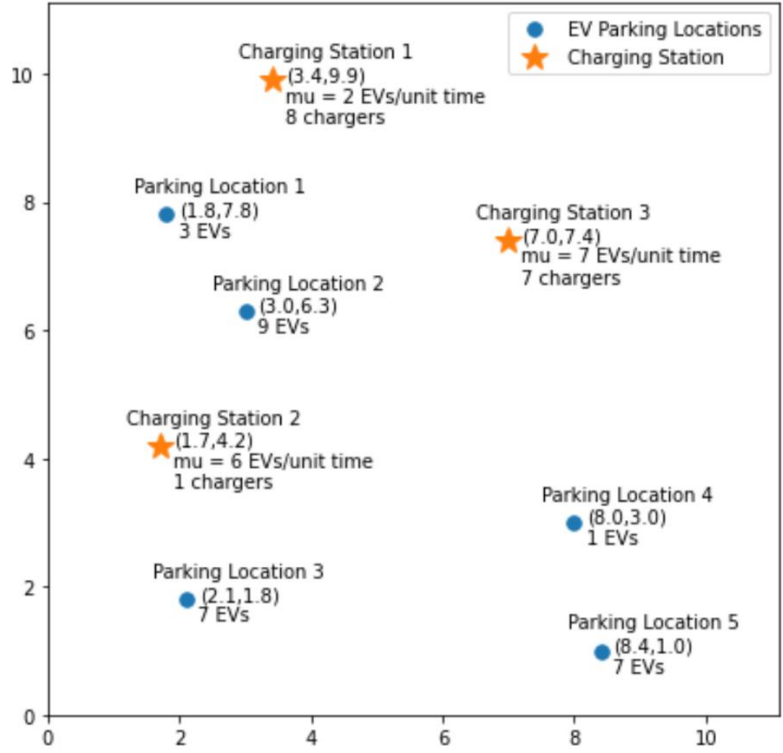

**Figure 2.** EV Locations and Charging Stations for Illustration

We run the algorithm with the above settings. The algorithm converged after 36,869 iterations for a tolerance of $\Delta = 10^{-4}$. The final assignment is shown in Eq. (9-10) with the final cost matrix in Eq. (11). Obviously, the result satisfies Wardrop's principle: for the EVs at the same parking location, the cost of all the charging stations that are used at user equilibrium are the smallest and equal. In this case, no one uses station 1 and few EVs use station 2. Only users from location 3 split their flows to access stations 2 and 3 with an identical equilibrium access cost of 0.52 for both stations.

$$X^n = \begin{bmatrix} 0 & 0 & 1 \\ 0 & 0 & 1 \\ 0 & 0.63 & 0.37 \\ 0 & 0 & 1 \\ 0 & 0 & 1 \end{bmatrix} \tag{9}$$

$$Flow\ Assignment\ Matrix = \begin{bmatrix} 0 & 0 & 3 \\ 0 & 0 & 9 \\ 0 & 4.39 & 2.61 \\ 0 & 0 & 1 \\ 0 & 0 & 7 \end{bmatrix} \tag{10}$$

$$T_{ij}^{n+1} = \begin{bmatrix} 0.63 & 0.57 & 0.4 \\ 0.68 & 0.52 & 0.35 \\ 0.91 & 0.52 & 0.52 \\ 0.91 & 0.72 & 0.37 \\ 1.01 & 0.77 & 0.47 \end{bmatrix} \quad (11)$$

## 4 NEW YORK CASE STUDY

### 4.1 Base Scenario

*4.1.1 Current EV Charging Station Configuration*

The charging location configuration on July 8, 2020, is used as the Base Scenario. The Base Scenario configuration of EV charging stations is shown in Figure 3(a). Using the Google maps data provided by DCAS, we only consider Level 2 chargers (ChargePoint, DOT lots, Nissan) and DCFCs. These charging stations are owned by DCAS and only used by the EV fleet owned by NYC. Level 1 charging stations are not considered, since the time it takes to fully charge is prohibitive for everyday use (around 10 hours to fully recharge). We exclude the 87 solar carports from the analysis since they have very low capacity. There are 567 total chargers, which consist of 563 Level 2 chargers and 4 DCFCs. These combine into 263 charging stations, among which 4 are DCFC stations and the rest are Level 2 charging stations. The numbers of chargers are shown in Figure 3(b). Charging stations with more chargers are distributed mainly on the north part of the city. The histogram of the number of chargers per station is shown in Figure 4. The charging station with the most chargers has 19 chargers, while 157 out of 263 charging stations have only 1 charger. For assignment, we assume that the charging rate of Level 2 chargers is 4 h/veh, and the charging rate of DCFCs is 0.5 h/veh. This is consistent with the 3-8 hr range of Level 2 chargers (Evocharge, 2021) and 0.5-0.75 hr range of DCFC (JD Power, 2021) (and similar values are used in Jung and Chow (2019) and Pantelidis et al. (2021).

*4.1.2 EV Parking Locations*

The EV parking location data that is shared from DCAS and aggregated to TAZs for privacy consideration. The 1,622 TAZs are from New York Metropolitan Transportation Council (NYMTC) used for the Year 2000 Best Practice Model (NYMTC, 2020). We assume that the centroids of the TAZs are the parking locations shown in Figure 5. There are 1,484 EVs owned by NYC, distributed over 512 of the TAZs. Hence, we have $|N| = 512$ demand nodes. Home charging is not considered in this case. We assume that all the EVs go to city-owned charging stations to recharge. There are two reasons. First, the EV fleet in the case study is owned by the city, which means that they are most likely charged at city-owned charging stations. Second, the unobservable home charging is accounted for through the calibration of the arrival rates to get results that match observed utilization rates (e.g. if there's a lot more home charging occurring, observed utilization would be lower with less visits per day). For the accuracy of M/D/C approximation, as shown in Figure 4, most of the charging stations have only 1 charger ($C = 1$), which is shown in Figure 1 to have low error. The approximation is accurate enough to be applied.

*4.1.3 Base scenario calibration*

For the base scenario, we know the locations of EVs but we do not know the arrival rate to charging stations. We calibrated the arrival rates (vehicles/day) by multiplying a factor to the number of EVs per zone such that the equilibrium charging frequency matches closely to a 7.6% average utilization ratio (New York State Energy Research and Development Authority, 2017). The factor can be interpreted as the inverse of the average charging frequency (days/ charge). The relationship between average charging frequency and average utilization ratio is shown in Figure 6. The average charging frequency of the EV fleet in NYC is around 3 days/charge.

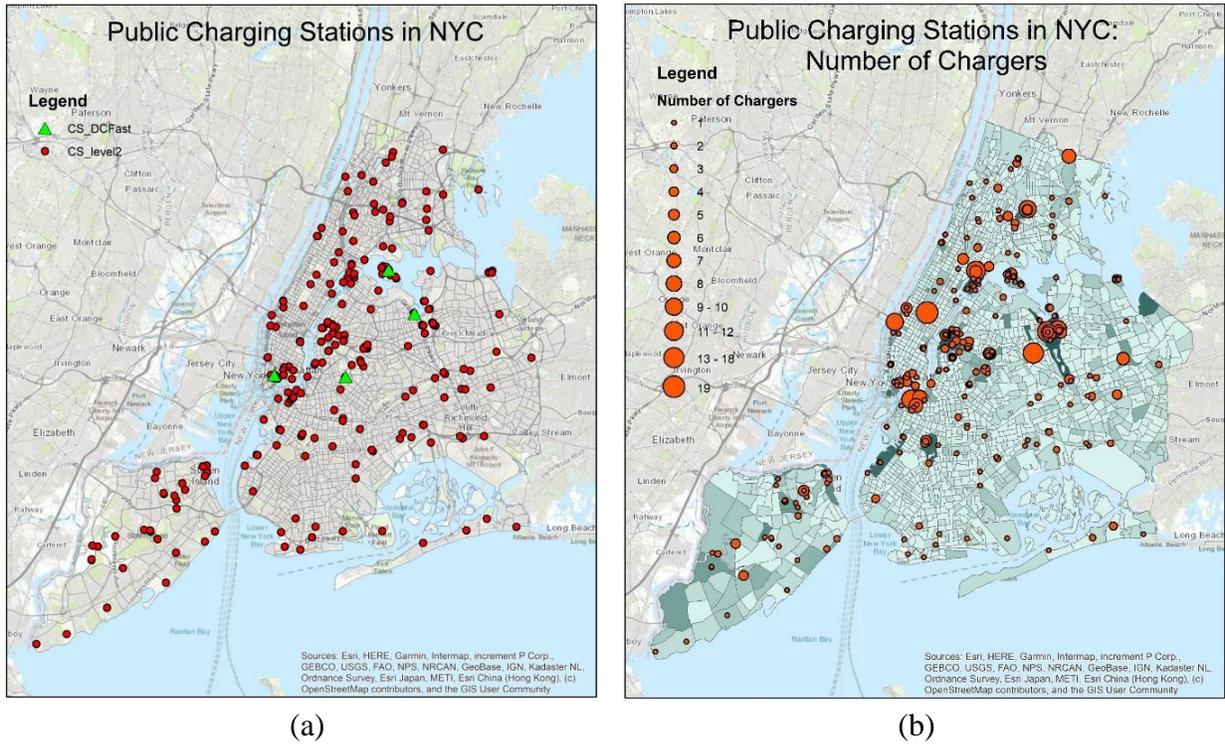

(a)                      (b)

**Figure 3.** (a) NYC Level 2 and DC Charging Stations and (b) number of chargers, as of July 8, 2020.

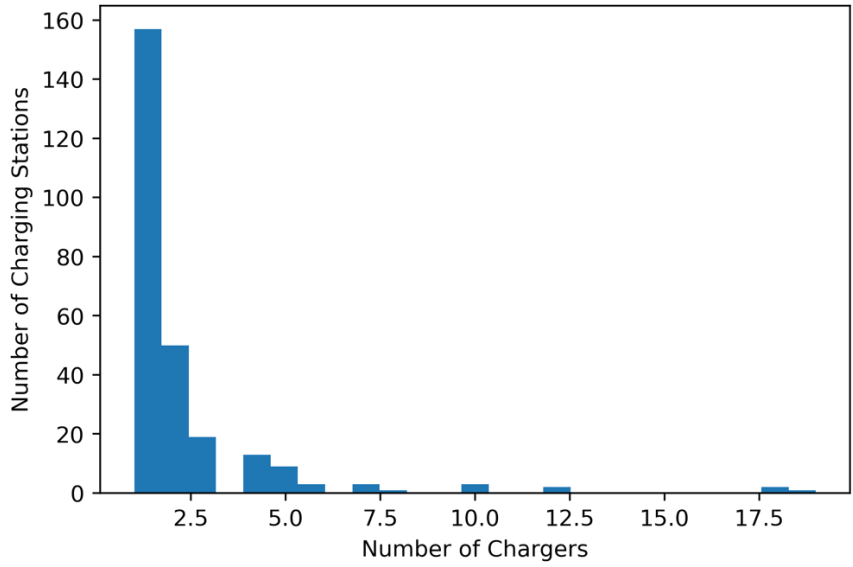

**Figure 4.** Histogram of Number of Chargers at each Charging Station.

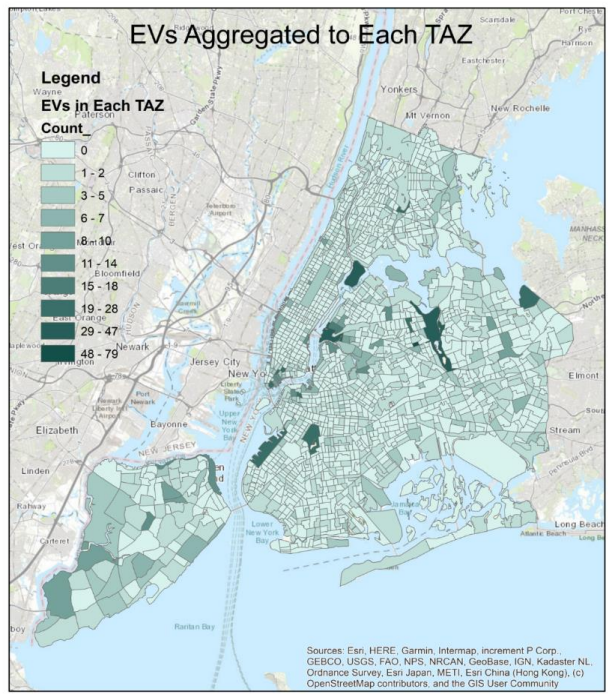

**Figure 5.** EV Parking Locations.

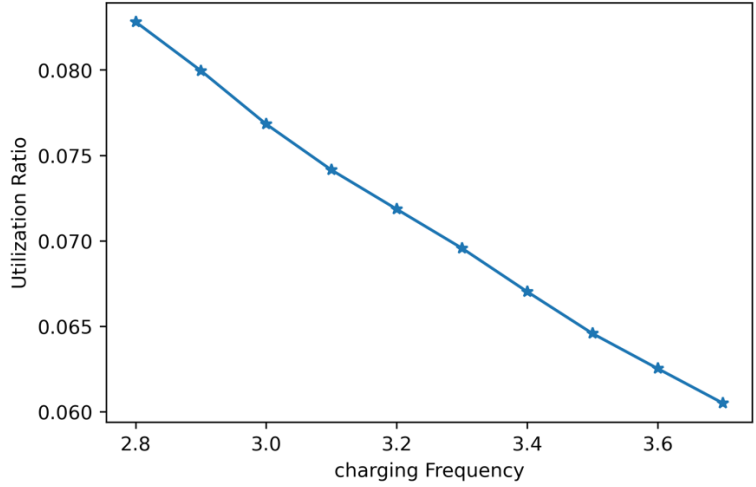

**Figure 6.** Relationship between Charging Frequency and Utilization Ratio for the Current NYC Charging Station Configuration (Jul. 8, 2020).

*4.1.4 Base scenario EV Assignment*

To perform the EV assignment for the Base Scenario, we assume that EVs travel at an average speed of 25 mph (citywide speed limit). We use ArcGIS to generate a shortest path between each TAZ centroid and charging station and compute the travel time correspondingly to form a travel time matrix. The weights of access time ($\omega_{acc}$) and charging time ($\omega_{char}$) are both set as 1 by default.

The proposed algorithm takes 1h 4min 8s (wall time with Intel® Core™ i5-7200U CPU and 8.00GB installed RAM, Python 3.7.2) and 12,623 iterations to converge to $\Delta = 10^{-3}$ for the Base Scenario. We summarize the assignment results by showing the range and average values of access time and the sum of access and charging time at each TAZ, as well as utilization ratio, expected queue delay and the sum of expected queue delay and charging time of each charging station in Table 3. At UE, total system access time is 27.53 days. Average access time for one EV is 80.15 min. The sum of system total access time and charging time is 83.87 days. The sum of average total access time and charging time is 4.07 hours per EV. This means that under the current charging location design, each EV takes 80.15 minutes (travel time and queue delay) to access a charger and 4.07 hours to access and charge a vehicle.

The utilization ratios and expected queue delays at DCFCs are significantly higher than Level 2 charging stations, which indicates that DCFCs are much more heavily used than Level 2 charging stations. The gap in the sum of queue delay and charging time between the DCFC and Level 2 charging stations is due to the cost of switching: going from DCFC to Level 2 would add a charging time of 3.5 hours, regardless the change in access time.

**Table 3**. Attributes at UE for Base Scenario

| TAZ | | Charging Station | | | | | | | | |
|---|---|---|---|---|---|---|---|---|---|---|
| Access Time (sum of travel time and queue | Total Time (min) | Utilization Ratio of Charging Stations | | | Expected Queue Delay (min) | | | Expected Queue Delay + Charging Time (hr) | | |
| | | Lvl 2 | DCFC | All | Lvl 2 | DCFC | All | Lvl 2 | DCFC | All |

|         | delay) (min) |       |        |        |        |      |       |       |       |       |       |
|---------|--------------|-------|--------|--------|--------|------|-------|-------|-------|-------|-------|
| **Average** | 73.5     | 245.3 | 6.36%  | 93.23% | 7.68%  | 2.6  | 206.6 | 5.7   | 4.043 | 3.944 | 4.042 |
| **Maximum** | 217.2    | 259.0 | 44.77% | 93.32% | 93.32% | 17.1 | 209.4 | 209.4 | 4.285 | 3.990 | 4.285 |
| **Minimum** | 1.1      | 235.7 | 0.00%+ | 93.14% | 0.00%+ | 0    | 203.6 | 0     | 4.000 | 3.893 | 3.893 |

As shown in Figure 7, long times to access and charge vehicles appear mostly in Queens, midtown Manhattan and Staten Island under the calibrated Base Scenario. The reason for the long access and charging times in these areas is the large queue delays caused by lack of charging stations, small numbers of chargers at charging stations, and the long charging time of Level 2 chargers. To alleviate the congestion at the charging stations and shorten the access time, one option is to add more chargers to current charging stations.

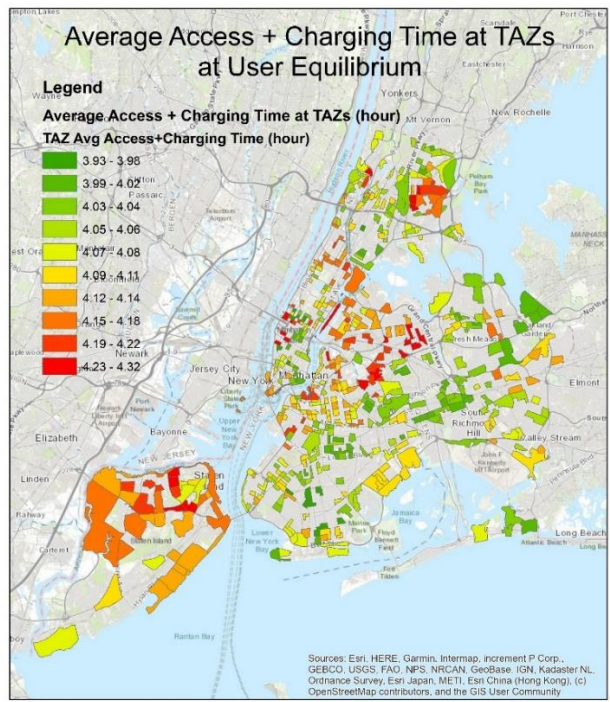

**Figure 7.** Average access time at each TAZ.

### 4.2 Evaluation of DCFC investment policies

We evaluate several growth policies to provide decision support for NYC. We consider adding DCFCs to the busiest Level 2 charging stations under different strategies and evaluate them with the proposed algorithm. Utilization ratio and queue delay are different metrics to identify busy charging stations and bottlenecks in the system. Hence, we compare two DCFC investment policies: adding to Level 2 stations with (1) the highest utilization ratio and (2) with the highest queue delay.

On the surface, both appear to be viable policies. However, the distribution of utilization ratios and expected queue delays of each charging station in the Base Scenario are different, as shown in Figure 8. This means that the "busiest" Level 2 charging stations selected according to queue delay and utilization ratio could be different. High queue delays are concentrated in areas with less or smaller-scale charging stations, which is mainly the south part of the city, while no significant geographic pattern can be found for utilization ratios.

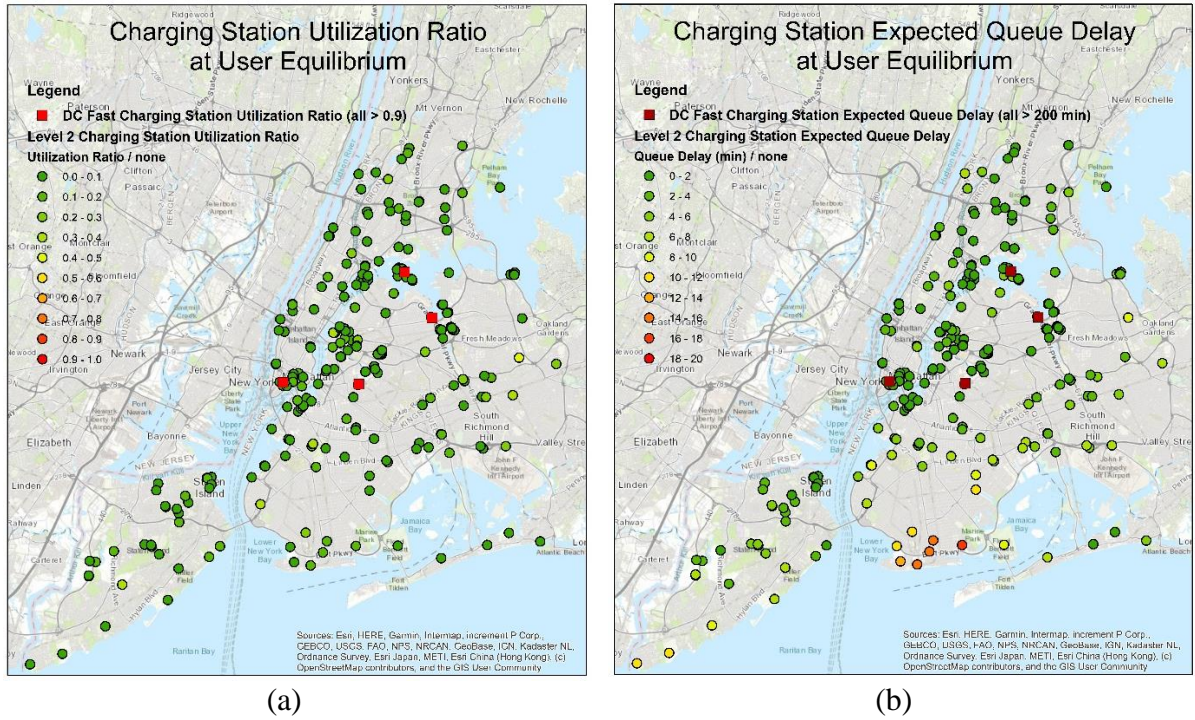

(a) (b)
**Figure 8.** (a) Utilization ratios and (b) expected queue delay at each charging station.

We design the following 10 scenarios shown in Table 3, where the first 5 scenarios (A1 – A5) add DCFCs to Level 2 charging stations according to the rank of utilization ratio and the last 5 (B1 – B5) add DCFCs according to the rank of expected queue delay. We add one DCFC to each Level 2 charging station chosen. The 5 scenarios of the 2 strategies have the same number of DCFCs and Level 2 chargers in each row. The improvement scenarios are shown in Figure 9.

**Table 3.** Improvement Scenarios with Different Strategies

| Strategy A: Utilization Ratio | | | Strategy B: Expected Queue Delay | | |
|---|---|---|---|---|---|
| Scenario | Level 2 Charging Stations Chosen (ID) | Number of DCFCs added | Scenario | Level 2 Charging Stations Chosen (ID) | Number of DCFCs added |
| A1 | 122, 132, 26, 225, 112 | 5 | B1 | 13, 227, 244, 243, 65 | 5 |
| A2 | 122, 132, 26, 225, 112, 89, 73, 18, 9, 125 | 10 | B2 | 13, 227, 244, 243, 65, 9, 173, 71, 120, 40 | 10 |
| A3 | 122, 132, 26, 225, 112, 89, 73, 18, 9, 125, 195, 93, 92, 109, 204 | 15 | B3 | 13, 227, 244, 243, 65, 9, 173, 71, 120, 40, 254, 110, 238, 178, 174 | 15 |
| A4 | 122, 132, 26, 225, 112, 89, 73, 18, 9, 125, 195, 93, 92, 109, 204, 181, 81, 237, 87, 98 | 20 | B4 | 13, 227, 244, 243, 65, 9, 173, 71, 120, 40, 254, 110, 238, 178, 174, 171, 204, 177, 214, 216 | 20 |
| A5 | 122, 132, 26, 225, 112, 89, 73, 18, 9, 125, 195, 93, 92, | 25 | B5 | 13, 227, 244, 243, 65, 9, 173, 71, 120, 40, 254, 110, 238, | 25 |

| | 109, 204, 181, 81, 237, 87, 98, 94, 144, 116, 202, 90 | | 178, 174, 171, 204, 177, 214, 216, 189, 181, 246, 23, 239 | |

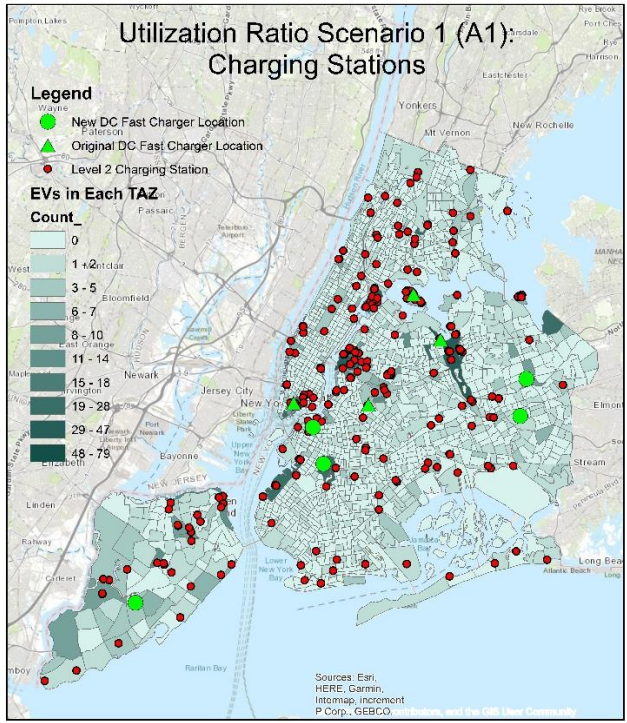

(a)

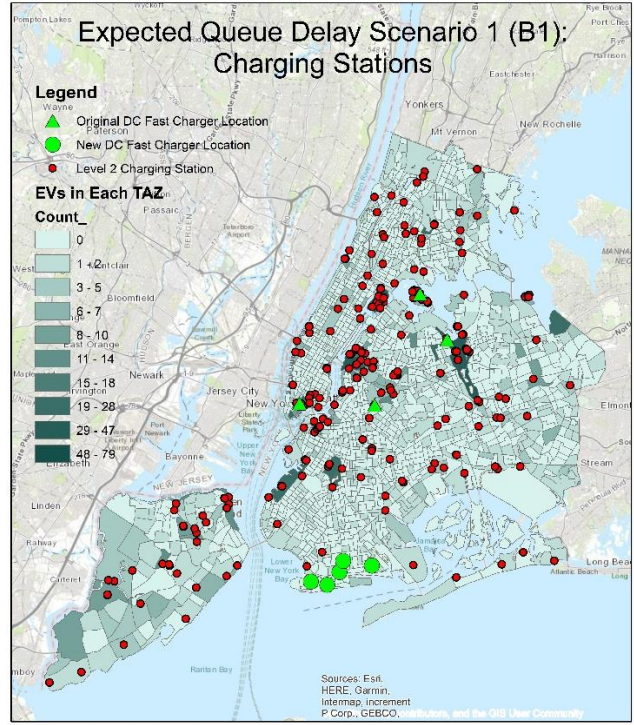

(b)

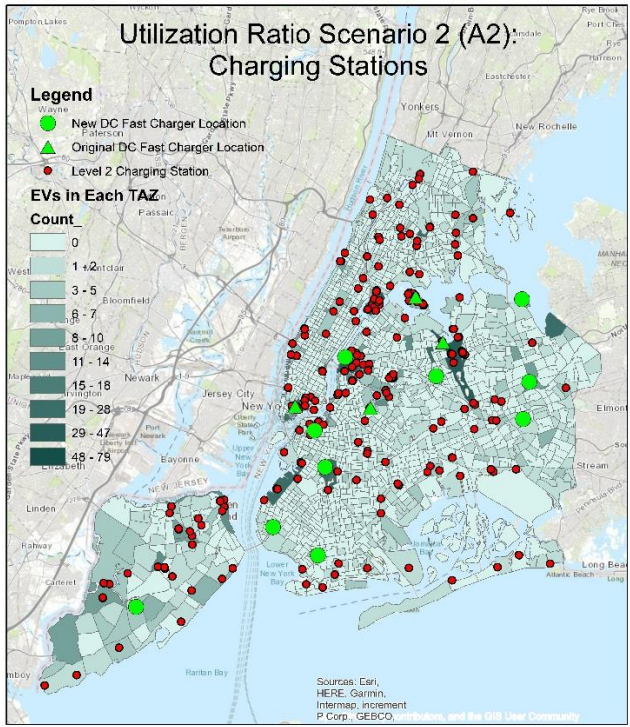

(c)

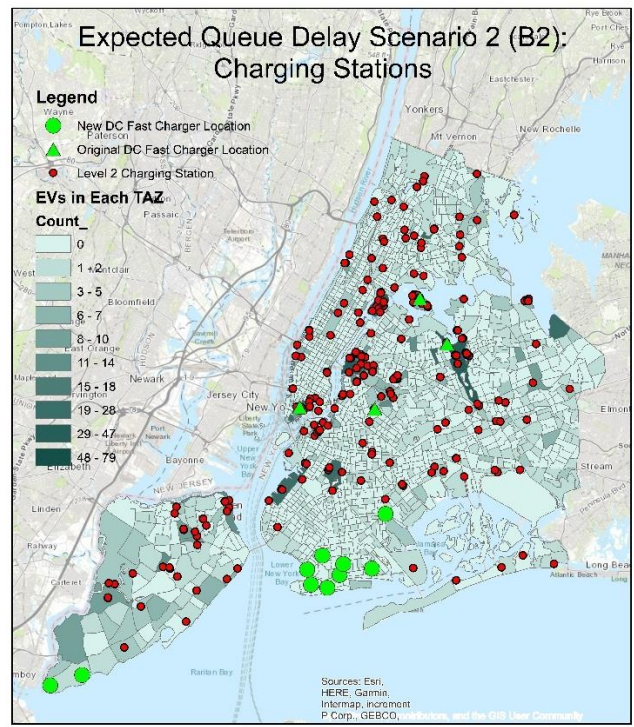

(d)

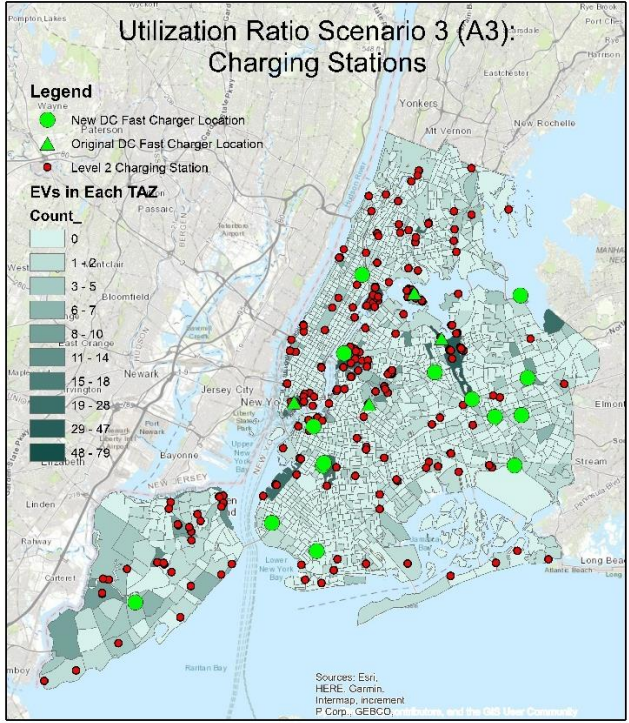

(e)

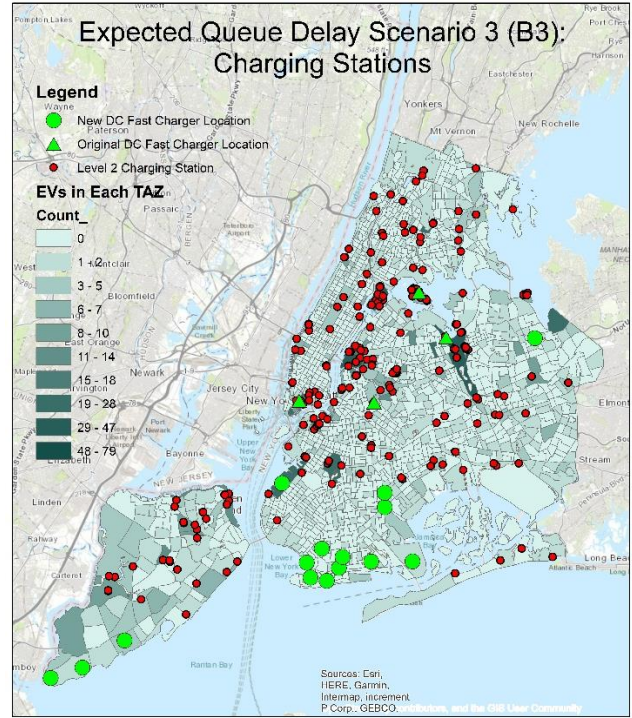

(f)

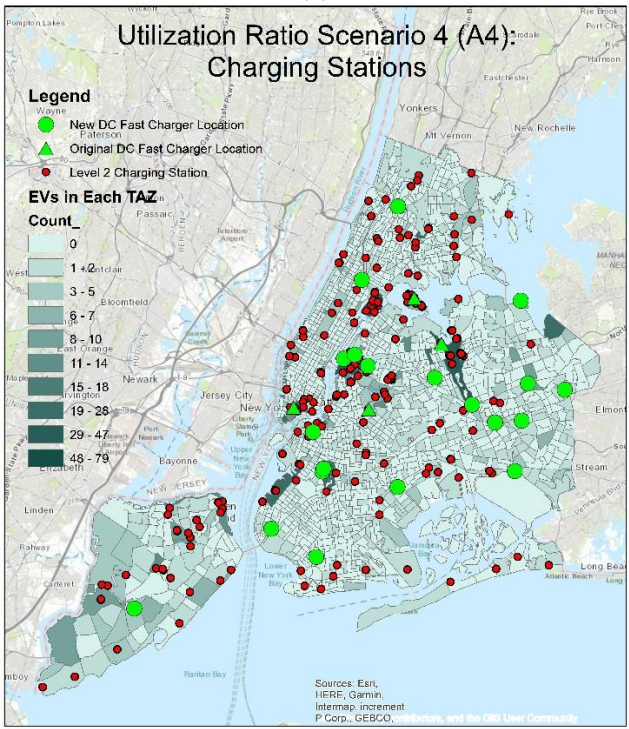

(g)

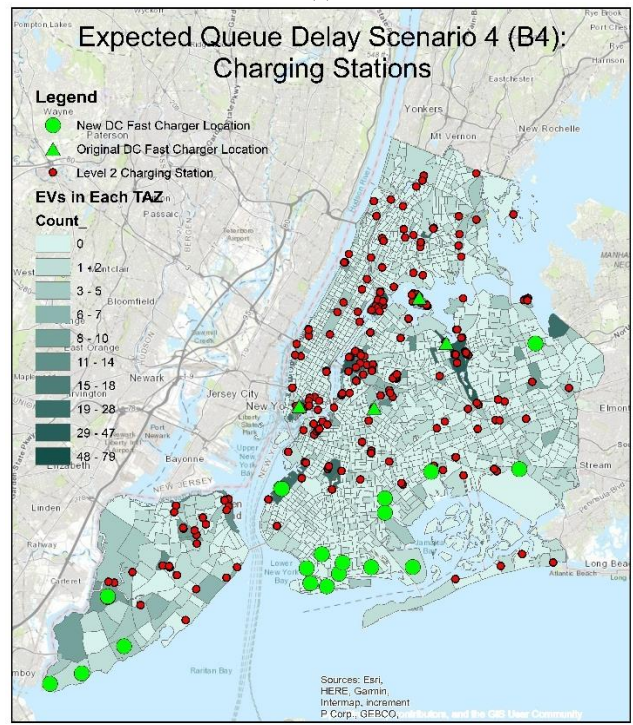

(h)

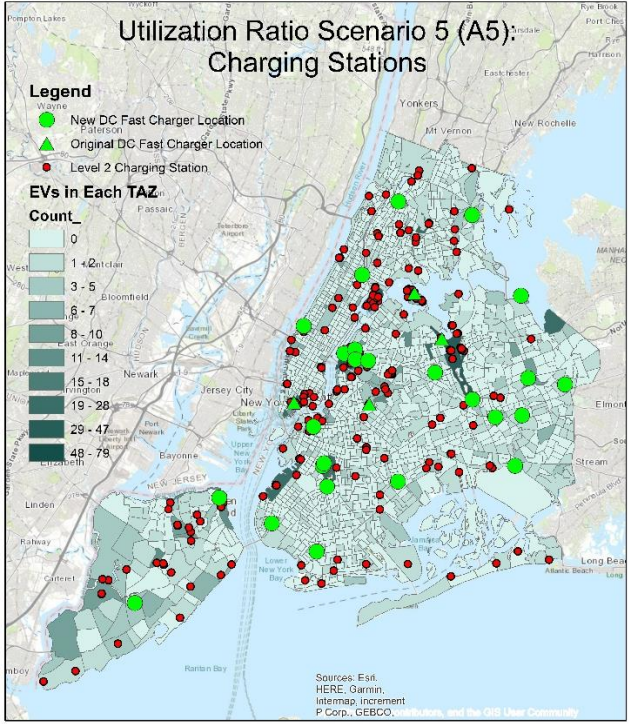
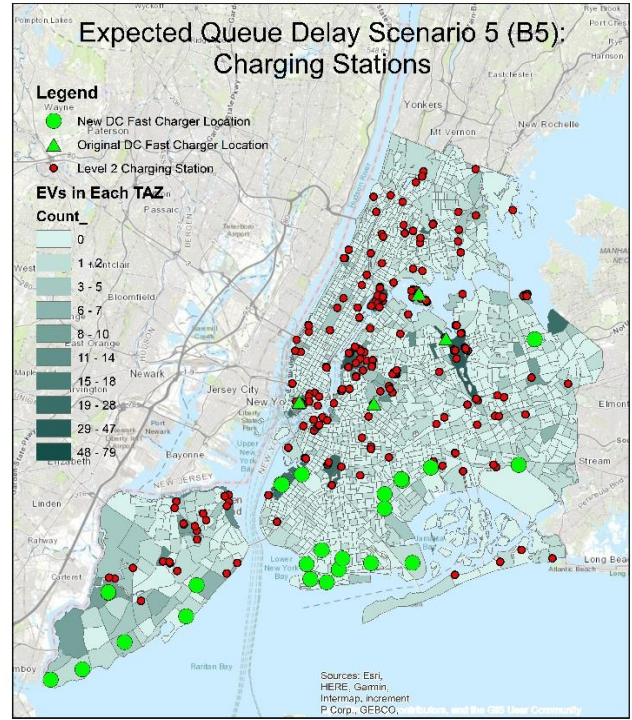

(i)  (j)

**Figure 9.** Charging station locations for (a) A1, (b) B1, (c) A2, (d) B2, (e) A3, (f) B3, (g) A4, (h) B4, (i) A5, and (j) B5.

Similar to the Base Scenario, we run the 10 scenarios and compute system total access time, the sum of system total access time and charging time, average access time, and the sum of average access time and charging time at UE for each scenario. Figure 10 and Table 4 shows how these 4 values change with DCFCs added, in which the 2 lines represent different strategies of selecting new DCFC locations. The starting point of the lines are the values from the Base Scenario.

**Table 4.** Different system time measures at UE for each scenario

| Scenario | A1 | A2 | A3 | A4 | A5 | B1 | B2 | B3 | B4 | B5 |
|---|---|---|---|---|---|---|---|---|---|---|
| System total access time (day) | 58.08 | 17.23 | 8.93 | 6.14 | 4.67 | 56.96 | 24.28 | 14.54 | 11.68 | 10.26 |
| System total access time + charging time (day) | 81.92 | 27.57 | 19.28 | 16.49 | 15.01 | 81.04 | 34.61 | 24.87 | 22.01 | 20.58 |
| Average access time (min) | 169.07 | 50.15 | 26.00 | 17.86 | 13.60 | 165.80 | 70.69 | 42.32 | 34.00 | 29.87 |
| Average access time + charging time (hour) | 3.97 | 1.34 | 0.94 | 0.80 | 0.73 | 3.93 | 1.68 | 1.21 | 1.07 | 1.00 |

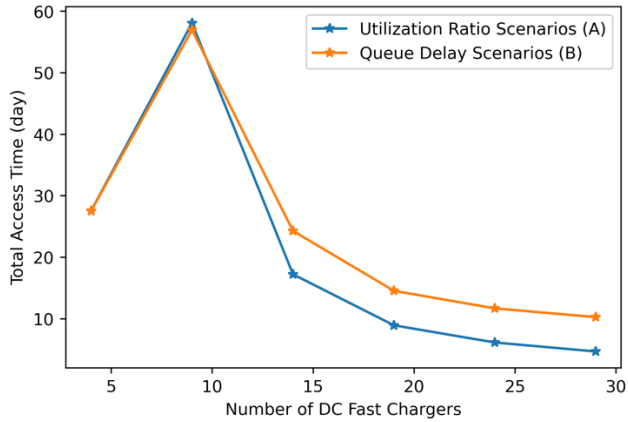
(a) Total System Access Time Change

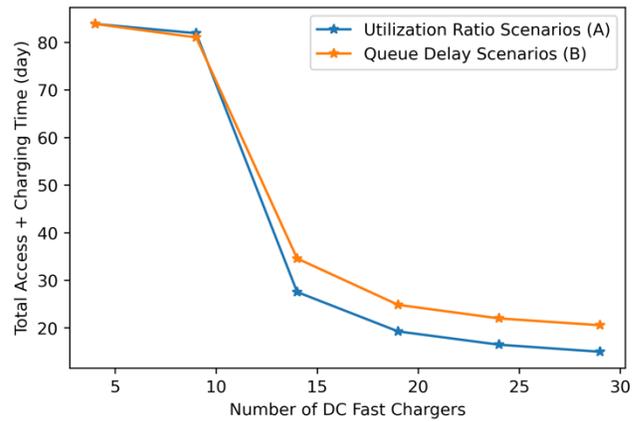
(b) Sum of Total System Access Time and Charging Time Change

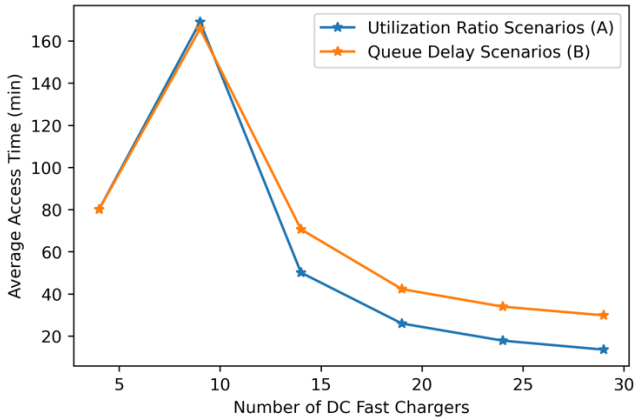
(c) Average Access Time Change

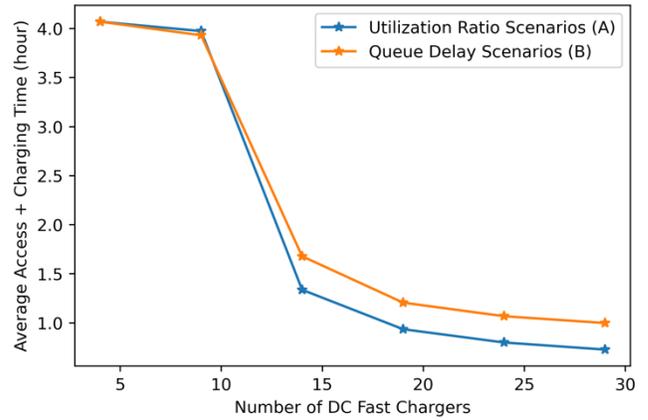
(d) Sum of Average Access Time and Charging Time Change

**Figure 10**. System changes caused by adding DCFCs.

For both strategies, we observe that the sum of system total/average access time and charging time monotonically decreases while the number of DCFCs increases, while system total/average access time increases after adding 5 DCFCs and then decreases after more are added. This is because when the number of DCFCs is small, they have significantly larger queue delay than Level 2 charging stations. Adding DCFCs attracts more EVs to DCFCs, which initially leads to higher system total/average access time. However, after more DCFCs are added, system total/average access times decrease due to the improvement of overall charging capacity of the system.

We observe that the reduction in system total/average access time with the increase in the number of DCFCs is not linear, nor is the sum of system total/average access time and charging time. There are significant drops between Scenario A1 and A2 (utilization ratio strategy) and Scenario B1 and B2 (queue delay strategy), which indicate that there are critical points where the increase in the number of DCFCs significantly improve the system efficiency.

### 4.3 Policy implications

The 2 different strategies result in different levels of change in system total/average access time as well as the sum of system total/average access time and charging time. Even though adding small numbers of DCFCs leads to similar system performance(A1, B1), adding more DCFCs with the utilization ratio strategy leads to more decrease in system total/average access time as well as sum of system total/average access time and charging time compared with expected queue delay strategy (A2, B2, A3, B3, A4, B4, A5, B5).

As a result, **utilization ratio** is a better criterion for identifying critical charging stations. In addition, the expected queue delay strategy tends to identify charging stations with small numbers of chargers. The reason is that with the same utilization ratio, charging stations with fewer chargers have longer queue delays. This leads to the tendency of identifying small-scale charging stations at low-demand areas, which also demonstrates the rationality of the utilization ratio strategy compared to the queue delay strategy.

The policy implication of this finding for charging infrastructure agencies is to consider adding new chargers to improve utilization ratio as opposed to reducing queue delay, as the former strategy proves to be more effective. The optimal number will depend on external factors related to budget and availability of space. It can be further aided by observing the shape of the curves in Figure 10.

Since no other queueing-based assignment study has been conducted with any empirical data, no comparison of these findings can be made with the literature.

## 5 CONCLUSION

Despite the dependency of EV fleets on charging station availability, the number of charging stations remains limited in many cities. Investment in additional charging infrastructure requires models to evaluate their performance. Three contributions are made in this study. First, having observed from the literature that people tend to prefer charging their EVs while at home, we propose an original EV-to-charging station access equilibrium assignment model. The equilibrium model considers users' choice in charging station with queue delay modeled as an M/D/C queue. This is the first such model in the literature and can prove useful to city agencies like NYC currently investing into charging infrastructure to achieve sustainability goals.

Second, the model is solved using a new derivative-free MSA algorithm that avoids having to compute either the objective value with the integral or the derivative in the step size determination, which can be problematic due to the M/D/C approximation function. Computational tests with a replicable toy network show that the model does indeed converge to a UE solution where no user can unilaterally change their station choice and be better off. The algorithm code is publicly available at our GitHub repository: https://github.com/BUILTNYU/EV_Charging_Station_Access_Equlibrium_Model.

Third, we use real location data shared by NYC DCAS of a fleet of 1,484 NYC-owned EVs located in 512 TAZs and 263 current Level 2 and DCFC stations to evaluate a Base Scenario for NYC's charging station investments for their city-owned EV fleet. The model converged to an equilibrium with $\Delta = 10^{-3}$ tolerance after 1h 4min 8s (wall time). The results suggest that much of the queue delays for users is incurred in Queens, midtown Manhattan, and Staten Island. Two DCFC investment strategies are tested to demonstrate the use of the model: a conversion of Level 2 stations with top utilization ratios versus those with top queue delays. The results imply that an investment policy based on utilization ratio would be much more effective in reducing overall access time.

Further work could be done collecting and modeling the distribution of charging time, as well as incorporating that into the assignment model to calibrate it. Although proven effective, the MSA algorithm converges slowly with large numbers of demand and charging nodes, especially when demand far exceeds charging supply. The convergence can be sped up using alternative algorithms like the Method of Successive Weighted Averages (Liu et al., 2009). With faster convergence, the assignment algorithm can be made into a web tool to evaluate different demand distribution and charging station configurations. This is an ongoing objective of the NYU VIP team on "Clean Fleets" working with NYC DCAS. In addition, while charging prices is not considered in this study, the model formulation can be trivially modified to take charging prices into consideration by constructing a generalized cost function which includes both time cost (converted to dollars using value of time) and charging prices.

With our analysis on the improvement scenarios, further work could consider facility location optimization as a bilevel problem. When NYC DCAS releases updates to their DCFC location plan, this tool can be used to help them analyze the performance, identify bottlenecks, as well as estimate optimal number of chargers to be added in a certain region given demand and budget.

**ACKNOWLEDGMENTS**
This research was supported by the C2SMART University Transportation Center. Stephanie Tam is an undergraduate member of the Vertically Integrated Projects (VIP) "Clean Fleets" team at NYU (https://wp.nyu.edu/vip/nyc-clean-fleet/) that is working with NYC DCAS to develop the web-based equilibrium model. The data shared by DCAS is gratefully acknowledged. Any errors and views expressed in this study solely belong to the authors.

**AUTHOR CONTRIBUTION STATEMENT**
The authors confirm contribution to the paper as follows: study conception and design: J.Y.J. Chow, T.P.Pantelidis, B. Liu; data preparation: S. Tam, T.P. Pantelidis, B. Liu; analysis and interpretation of results: B. Liu, J.Y.J. Chow, T.P. Pantelidis; draft manuscript preparation: B. Liu, T.P. Pantelidis, J.Y.J. Chow, S. Tam. All authors reviewed the results and approved the final version of the manuscript.